\documentclass[10pt, oneside]{article}   	% use "amsart" instead of "article" for AMSLaTeX format
\usepackage{geometry}                		% See geometry.pdf to learn the layout options. There are lots.
\usepackage{graphicx,bm,color}				% Use pdf, png, jpg, or eps§ with pdflatex; use eps in DVI mode
\usepackage{amssymb,hyperref,multicol}
\usepackage{rotating}

\title{A discussion of the cross section $\bar\nu_e+p\to  e^+ + n$}
\author{G.~Ricciardi$^1$, N.~Vignaroli$^{1,2}$, F.~Vissani$^3$,\\
\small $^1$Dipartimento di Fisica  E. Pancini,   
Naples U.  Federico II and INFN Naples, Italy\\[-0.3ex]
\small $^2$Dipartimento di Matematica e Fisica, Salento U., and INFN 
Lecce, Italy\\[-0.3ex]
\small $^3$INFN Laboratori Nazionali del Gran Sasso,
 L'Aquila, Italy}
\date{}							% Activate to display a given date or no date

\begin{document}

\renewcommand{\abstractname}{\vspace{-7mm}} %-\baselineskip}}
\maketitle
%\section{}
%\subsection{}
\begin{abstract}
We discuss the interaction cross section $\bar\nu_e-p$ due to charged currents, of major importance for neutrino detection.
We present the history of its understanding and  highlight the aspects necessary for its precise evaluation. 
We examine the three most recent determinations and, on the basis of the most recent one, tabulate its updated values and assess its uncertainty.
\end{abstract}

\section{Importance of the IBD cross section}\label{sec:i}
The process 
$\bar\nu_e + p \to e^+ + n 
$
has been the first by which neutrinos were directly observed \cite{rc}.
While it can be correctly described as a 
reaction between electron antineutrinos and protons, it is often called `inverse beta decay' because,  
in the context of quantised field theory, it shares the same amplitude as the beta decay of the neutron.
 It is widely used in 
%Over time, it has demonstrated 
% for electron antineutrino detection 
%%it has shown its applicability 
%it has been shown how it occurs
%in various situations,
water- or hydrocarbon-based detectors, which are relatively cheap materials 
and  rich in target protons. 
 Future experiments for detecting antineutrinos from 
 reactor and from gravitationally collapsing supernovae,  see e.g., \cite{db, ju,hk}, 
 will collect very large samples of such events and will require the cross section to be known precisely.
This consideration alone has motivated a long-standing interest in its theoretical estimation.  
In this contribution, we first point out the significant elements for calculating it (section~\ref{sec:h}), and we do so 
 in an entertaining way, that is, by retracing in broad outline 
the interesting history of its theoretical understanding. In the second part, after presenting 
the three most modern and accurate calculations (section~\ref{sec:c}),  we  focus on the more recent one. We discuss the 
most reliable expression of the cross section 
by estimating what the residual uncertainties are (section~\ref{sec:u}). The last part 
(section~\ref{sec:s})  is devoted to a brief overview of the current status 
and  future prospects.

\section{Brief history of the IBD reaction}\label{sec:h}
In order to highlight what ideas underlie the description of the IBD cross section, and to do so in an agile manner, we take a cursory look at its history. Before it was possible to speak of a cross section, the concept of the neutrino itself had to be developed; then, in about 20 years, scientists moved from the first Hamiltonian theory of beta rays 
 to the modern description of interactions (V-A theory). Since then numerous other advances have occurred, some of which are particularly relevant to the quantitative discussion of the cross section: the understanding of the Cabibbo angle and an adequate description of hadronic interactions.

%\section{The IBD cross section}

\paragraph{Evolution of the idea of neutrino}
Let us begin presenting the different ideas of the neutrino, formulated around 1930s:\\
{\em 1. Pauli 1930} \cite{pauli}
introduces the neutrino as a constituent of the atomic nucleus and assumed that this particle is emitted in $\beta$ decay\footnote{This model has no relativistic characteristics and in particular has no connection with Dirac  idea of antimatter.}.\\
{\em 2. Fermi 1933-1934} \cite{rs33} describes neutrinos as 
relativistic (Dirac) fermions, completely analogous to the electron. Due to the chosen formalism  \cite{fermos} - see also below - antineutrinos and  neutrinos are different\footnote{Fermi's neutrino concept corresponds  
to what is now called the `Dirac neutrino'. This term is widespread today, but Fermi does not use it and we do not know
any work of Dirac describing such a neutrino concept.}.\\
{\em 3. Majorana 1937} neutrino idea \cite{maj}  consists of the assumption that the neutrino and the antineutrino are the same particle. \\
{\em 4. Weyl's 1929} relativistic wave equation \cite{weyl} is simpler than Dirac's one and describes electrons with zero mass. 
Its relevance  to neutrinos will become apparent much later\footnote{Weyl's hamiltonian is $H=\pm \vec{\sigma}\, \vec{p} c$; 
 at the time it was set aside 
because of the peculiar coupling between  
 the momentum $\vec{p}$ and the spin 
 $\vec{S}=\hbar \vec{\sigma}/2$ - polar and axial vectors - which was completely outside the accepted patterns.}.\\
%
%{\em Weyl's 1929}  relativistic wave equation $H=\pm \vec{\sigma}\vec{p} c$ is simpler than Dirac's and describes only particles with zero mass; its relevance for neutrinos will emerge later\footnote{At the time it was shelved as it contains a `strange' coupling between spin $\vec{\sigma}$ and momentum $\vec{p}$.}.
As we will see in the next paragraph, the first discussion of the IBD cross section relies on the second type of concept. The  importance of the difference between neutrinos and antineutrinos will emerge later, and even later will be understood that it is possible to define, through this difference, a conserved lepton number. 
What about the   $3^{rd}$ and $4^{th}$ concepts? 
Majorana's proposal is explored and temporarily shelved. 
In the mid-1950s, Weyl's formalism was recognised as valid for the description of neutrinos; its compatibility with Majorana's proposal, accepted today, would only be understood later, slowly and with some hesitation \cite{mois}.
%\footnote{In fact, some uncertainty or doubt in this regard still persists.}.

For completeness, we mention here another important and later evolution in the comprehension of neutrinos, 
that does not concern us directly: 
this occurred after  it was realised that  there are more types of neutrino.  In 1962, Sakata and collaborators proposed that the neutrinos that interact with the charged leptons, via weak interactions, are not necessarily mass eigenstates, but possibly superpositions of mass eigenstates. This framework allowed Pontecorvo to reformulate the proposal of 1957 into its modern form, describing what we currently call {\em neutrino oscillations.}
%pontecorvo e PMNS

\begin{figure}[t]
\centerline{\includegraphics[width=0.95\textwidth]{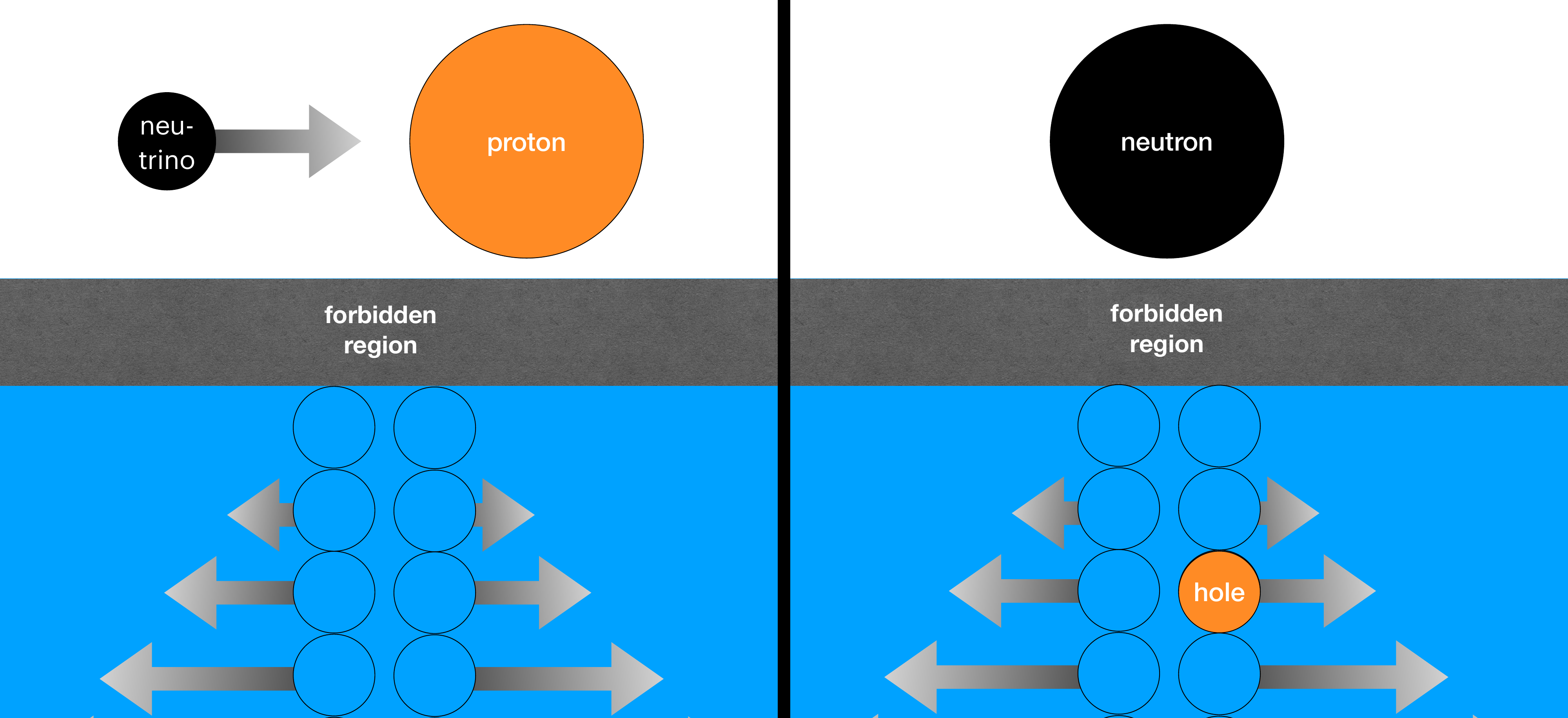}}
%\vskip4mm
%\centerline{\includegraphics[width=0.95\textwidth]{beta2s.pdf}}
\caption{\small\it Description of  IBD cross section 
 in the formalism of second quantization. 
 %emphasising the role of the   Dirac sea of electrons.  
{\em Left panel:} Initial state of the process; the neutrino hits the proton, the 
electron states of Dirac's sea are all occupied. {\em Right panel:}
Final state of the process. The nucleon changed its isospin state and became a neutron; the hole formed in the Dirac sea can be thought of as an anti-electron (=Dirac hole theory).\label{fig1}}
\end{figure}

\paragraph{From the theory of beta rays to the V-A interaction}
The first  theory of Fermi of $\beta$ decay, dating 1933 \cite{rs33}, introduces a constant $g$ which carries   dimensions of energy$\times$volume, namely, it is an inverse of a square mass in natural units. This enters the interaction hamiltonian
$H  = g\, \bm{\tau}_+\, \bm{\Psi}^{\dagger}\delta\bm{\Phi}^* + \mathrm{h.c.}$
 that allows the conversion of a neutron into a proton (the adimensional  isospin operator $\bm{\tau}_+$)
 and the appearance of a neutral and a charged lepton 
described by the fields $\bm{\Phi}=\sum_\sigma \phi_\sigma \bm{a}_\sigma$ and  $\bm{\Psi}=\sum_s \psi_s \bm{a}_s$, summed over positive {\em and} negative energies 
 (that form the leptonic current
 $\bm{J}_-= \bm{\Psi}^{\dagger}\delta\bm{\Phi}^* $). 
This theory relies heavily on the old procedure of quantisation (second quantization),  based on the existence of Dirac sea of electrons and of neutrinos, but it allows a lot of 
useful inferences; see e.g.~\cite{wick} and figure~\ref{fig1} for the description of the IBD cross section in this formalism.

In 1934, Bethe and Peierls \cite{bp}
observed that  the ratio of two quantities $\Gamma_n=\hbar/\tau_n$ and $\sigma_{\bar\nu_e p}$  
could be estimated roughly by 
dimensional considerations\footnote{Their estimate  is $ \sigma_{\bar\nu_e p} \sim 
\ell^3/(c \tau_n)$ where  $\ell=\hbar/(\mu c)$ and %; $c$ is the speed of light, $\hbar$ the reduced Planck constant and 
$\mu$ a mass characteristic of the IBD cross section,
as the mass difference between neutrons and protons or the mass of the electron.}: thus, from the measurement of $\tau_n$ one could get an idea of the size of  
the IBD cross section.
Since $\sigma_{\bar\nu_e p}$ turns out to be very small, this argument 
indicated that it was very difficult to see the antineutrino. This observation inclined Bethe and Peierls toward a  
pessimistic attitude \cite{bp} and they concluded that: 
{\sf ``there is no practically possible way of observing the neutrino''}. 
But 20 years later, in 1956,  Reines and Cowan \cite{rc}, 
succeeded in revealing the effects of this reaction. This was achieved by using detectors containing large masses of protons and exposed to copious flux of antineutrinos,  emitted by the first nuclear reactors ever built.
The existence of antineutrinos had been conclusively demonstrated, as eventually recognised by the 1995 Nobel Prize in Physics awarded to Reines; moreover IBD cross section could be measured. 

Meanwhile, the new fermion quantisation procedure proposed by Majorana \cite{maj} had become popular, making it no longer necessary to assume the existence of the Dirac sea.
However,  at that moment, Majorana's hypothesis on neutrinos \cite{maj} apparently lost its attractiveness, after a null result obtained by Davis~\cite{dacu} shortly before Reines and Cowan measurement\footnote{Davis tested that the particle produced in the reactors did not trigger events from $\bar\nu+{}^{37}{\rm Cl} \to {}^{37}\rm{Ar}+e^-$ \cite{dacu}. See \cite{mois} for a review of a detailed account of the subsequent phase of the discussion.}.

The last step of this story begins with the realisation that parity is not respected in weak interactions \cite{ly1,wu}, which harmonises well with the idea that the neutrinos are described by Weyl's equation  \cite{abd,dau,ly2} and with the ensuing conclusion that weak interactions have a V-A (vector-axial) structure \cite{sud,fg} also called chiral.  These hypotheses implies that neutrinos 
 are distinguished from antineutrinos by their helicity: Goldhaber's tests
\cite{gold}  confirmed the validity of this conclusion.

At this point we are close to the modern understanding; but the new positions concerning the interactions 
required to reconsider the conclusions drawn from Davis's experiment \cite{dacu}. Indeed, in the limit that usually counts in the laboratory, the ultra-relativistic limit,  
the V-A structure conceals the effects of the neutrino mass, including possible leptonic number violations caused by the Majorana mass.
Therefore, Davis' null result supports the view that neutrinos and antineutrinos are different when $E\gg m c^2$, 
but it is of no use if we want to know what is the value of Majorana's neutrino mass $m$: very special experimental situations are needed to probe it\footnote{See again~\cite{mois}; but for our purposes, it suffices to note that the search for Majorana's neutrino mass effects involves an ever increasing number of experiments.}.

Let us summarise: upgrading 
Fermi's Hamiltonian  as a  modern quantum field theory and within 
V-A structure, neutrinos fit Weyl's ideas and do not contradict in any manner Majorana's hypothesis, which remains attractive for its own
 reasons, observational and theoretical:  see e.g.~\cite{mois,matteo}.

\paragraph{Further relevant progresses}
There have been many other important advances since that time: they concern 1)~the inclusion of radiative corrections, 2)~the description of the effects of hadronic interactions, 3)~the effects of hadron mixing, and finally 4)~the conceptualization of the quark model. We note that some effects of radiative corrections are already considered by Fermi in 1933 \cite{rs33}, and the available calculations \cite{Kurylov} are adequate for current needs; furthermore,  there is no actual need to  rely on quark model concepts to discuss the IBD cross section. Therefore, for the purposes of our discussion, we  focus on the second and third aspects.

As far as hadronic interactions are concerned, recall that Yukawa \cite{yuk} introduces a boson to describe, in a way similar to QED, Fermi's interactions. As it is well known, that boson was subsequently identified with the charged pion. In the 1950s it was realised that there are other possible mediators - generically called resonances\footnote{Subsequently these considerations will be connected with the QCD, i.e. with the theory of gluons and quarks, and the resonances will be thought of as bound states of the $u$-$\bar{d}$ quarks that interact with $W^+$.} - and some of them are linked to the electromagnetic interactions of the proton and neutron by CVC and PCAC. As far as we are concerned, the outcome is the following \cite{ll,ria}: the Lorentz invariant decomposition of the hadronic current includes momentum-dependent form factors for vector and for axial parts
$f_i(q^2)$ and $g_i(q^2)$, $i=1,2,3$, whose value in $q^2=0$ can be measured and whose evolution to some extent can be constrained.

The last point relevant to the IBD cross section concerns hadronic  mixing, signs of which had been seen since the late 1950s. As there were some heated discussions on the occasion of a past Nobel Prize, perhaps it is helpful to recall the papers that  are relevant for the point under discussion. The first one \cite{gl} postulates that the vector current {\em matrix elements} includes a mixing between neutrons and $\Lambda$-particles to explains certain observations. The second one \cite{mns} extends the proposal to the axial matrix elements, assumed to obey exactly V-A structure. The third  paper \cite{cab} has a much broader scope \cite{cern}: it explores the consequences of CVC and SU(3) flavor symmetry of hadrons  for the {\em operators} (currents) that cause weak interactions. In this way, mathematical predictions for numerous processes are obtained: the  matrix elements between hadrons are  shown to  depend on one angle,  the Cabibbo angle, and on another parameter characteristic of axial interactions. This applies, in particular, to the transition between neutrons and protons we are interested in.

%
%The last point relevant to the IBD cross section concerns hadronic {\em mixing}, signs of which had been seen since the late 1950s. As there were some heated discussions on the occasion of a past Nobel Prize, perhaps it is helpful to briefly recall the works that  are relevant for the point under discussion. The first \cite{gl} postulates a mixing between protons and $\Lambda$-particles for the vector current matrix elements to explain the observations. The second \cite{mns} extends the proposal also to the axial matrix elements, assumed to obey exactly V-A structure. The third \cite{cab} has a much broader scope: it explore the consequences of CVC and SU(3) flavor symmetry  for the currents that cause weak interactions and also for the hadrons. In this way, mathematical predictions for numerous processes were obtained: the matrix elements of the currents between hadrons are  shown to  depend on one angle,  the Cabibbo angle, and on another parameter characteristic of axial interactions. This applies, in particular, to the transition between neutrons and protons we are interested in.
%

\section{The IBD cross section}\label{sec:c}

\paragraph{The three most recent determinations} 

The first modern calculation of the IBD cross section dates back to 1999 and is due to Vogel and Beacom \cite{vb}. 
The authors  systematically assessed the
%supposedly small and previously neglected effects such as % alcuni lavori precedenti ne avevano gia' notato la rilevanza
effects  of nucleon recoil,  as well as weak magnetism, whose understanding
began in the  1930s %and ended  non si finisce mai..:)
and was  finalised by  Gell-Mann \cite{Gell-Mann:1958sai}. 
Vogel and Beacom showed that these  effects are rather relevant for the positron angular distribution at  the desired accuracy. 
%This paper 
They adopted
%shows an expansion of the result 
an expansion in powers of $E_\nu/M$, where $M=(m_n+m_p)/2$ is the average mass of the nucleon, 
reliable in the region below 
$E_\nu <60$ MeV; they
also offered several useful analytical results and discussed the pointing of supernovae through the IBD reaction.
%the application to the pointing of supernovae through the IBD reaction.  

Three years later, Strumia and Vissani \cite{sv} produced a fully relativistic calculation based on the 4 known form factors, virtually valid at all energies. The result compares very well with the one of the previous calculation, when all relevant terms are included, and the ease of implementation of the expression is comparable. 
This paper gives an estimate of the uncertainty: at lower energies it is 0.4\%, while at higher energies there is an additional error due to the uncertainties of the form factors whose effect is estimated to be 
0.4\% $\times (E_\nu/50\mbox{ MeV})^2$ for $E_\nu$ below about 200 MeV.

Two decades later, Ricciardi, Vignaroli and Vissani  \cite{rvv} 
%attempted to 
improved the assessment of uncertainty in expectations: they verified the insignificance of `second-class currents', updated the relevant parameter values and performed a number of checks. The first result is obtained by maximising the parameters, while taking phenomenological constraints into account; the others are discussed in the next paragraphs, together with the estimation of the uncertainty in the IBD cross section.

  \paragraph{Numerical table of the IBD cross section}
In the last part of this note, entirely based on Ref.~\cite{rvv}, we overview the current values and  
uncertainties on the IBD cross section. 
In this paper the analytical formula of the cross section is given. It is also tabulated using the input values
$V_{ud} =0.97427$, $\lambda=1.27601$ and $r_A^2=0.416$ fm$^2$. 
While the first two values are currently the best fit ones, the latter, which corresponds to $M_A=1060$ MeV, is not - even if it lies within the uncertainty range $r_A^2=0.46\pm 0.16$ fm$^2$ discussed below. 
Therefore, the table~\ref{tab:xsec} presents the  IBD cross section, calculated 
for  the set of values $V_{ud} =0.97427$, $\lambda=1.27601$ and 
$r_A^2=0.46$ fm$^2$.

\paragraph{Uncertainties}
\label{sec:u}
The  radiative corrections of QED  to the cross section are calculated at 
leading order and included. Next order corrections and other effects such as 
isospin breaking are estimated to be small. In short, for the accuracy of interest, the leading 
uncertainties are simply due to three input parameters, and more precisely:\newline
$\bullet$ at lowest energies, the Cabibbo angle  
and the axial coupling;\\
$\bullet$  at higher energies, the axial mass $M_A$, or better (as we will discuss) the axial radius $r_A$.\\
Let us discuss their values and uncertainty ranges.

\noindent{\em Low energy region:}
First of all, let us summarise the way we treat the two relevant parameters.

The 
mixing element $ V_{ud}=\cos \theta_C $, that multiplies the amplitude of transition, 
can be probed:\\
$\star$ Directly, from the super-allowed transitions (we use Ref.~\cite{Hardy:2020qwl});\\
$\star$   Indirectly, exploiting the unitarity of CKM matrix,  
using the  values of $V_{us}$ and $V_{ub}$ 
given in \cite{ParticleDataGroup:2020ssz}\\
The two results are not in perfect agreement; thus, we include the scale factor  
$S=\sqrt{\chi^2/(N-1)}=2.0$
 for a conservative estimation of the uncertainty.

%$\star $ 
 The axial coupling   $g_1(q^2)$ is usually 
 presented in terms of a ratio with the vector form factor  $f_1$ 
 at zero momentum transfer, namely the axial coupling
  $\lambda=-g_1(0)/f_1(0)$. There are 
  eight measurements with polarised neutron decay, and the 
most recent one \cite{Markisch:2018ndu} is very precise.
Czarnecki, Marciano and Sirlin \cite{Czarnecki:2018okw} suggested 
to omit pre-2002 values, on behalf of  potentially large correction factors not completely under
control;  we have preferred to include them, but enlarging their error by a
factor 2, which implies a larger $S$.
 %enlarging  the  $S=2$. 
The 
resulting  value $\lambda= 1.2760(5) $ is within $1\sigma$ from Ref.~\cite{Markisch:2018ndu} and agrees with the global average. 
 
 Now, once the range of these two parameters is known, 
 we have a {\em prediction} 
 for the neutron decay lifetime $\tau_n$. On the other hand, this quantity is measured; 
 so in principle this measurement could help us to improve the inferences on the IBD cross section.
 %we can check it for consistency. 
 However, there are two sets of measurements of $\tau_n$
 that are among them incompatible: the total lifetime measured using  trapped ultra-cold neutrons is found to be $\tau_n(\mbox{tot})=878.52 \pm 0.46$ s,
 but the value deduced   using beam neutron and measuring the decay products is about 10 seconds longer: 
 $\tau_n(\mbox{beam})=888.0 \pm 2.0$ s. The data of $V_{ud}$ and $\lambda$ 
 are perfectly consistent with the former value, and incompatible with the latter. 
 There is no simple theoretical way out; the first suspect becomes an unknown systematic error.
 Efforts should be made to understand the incompatibility between the two set of $\tau_n$ 
 measurements. 
 %The inclusion of the measured value of $\tau_n(\mbox{tot})$ would allow us to narrow further 
 %the ranges of $V_{ud}$ and $\lambda$, but we prefer not to do so and to proceed as conservatively as possible.  
 %, using $\tau_n(\mbox{tot})$  as a consistency check.
% We believe that  the ranges of   $V_{ud}$ and $\lambda$  are not called into question, 
% but  in order to be as conservative as possible, we prefer to use 
% the measurement of $\tau_n(\mbox{tot})$ only as a consistency check.
 %We believe that  the ranges of   $V_{ud}$ and $\lambda$  are not called into question, but 
%in order to be as conservative as possible, we prefer not to use 
 %$\tau_n(\mbox{tot})$ data to improve  the  expectation for the cross section. 
 
 In summary, by  propagating the uncertainty factors we find that the cross section is known with
 %within 
$\delta\sigma_{\bar\nu_e p}/\sigma_{\bar\nu_e p}= 0.1\% $
for low values 
 of electron anti-neutrino energies: this is 4 times
 better than in~\cite{sv}.
 
\noindent{\em High energy region:}
 Past determinations of  the cross section have used the value of the axial mass $M_A$,  
which is measured at energies  $E_\nu \sim $  GeV or above, assuming that 
 the axial form factor behaves as a double dipole,
 $g_1(q^2)/g_1(0)=1/(1-q^2/M_A^2)^2$.  This value is quite precise \cite{Bodek:2007ym}
 $M_A=1014\pm 14$ MeV and it is supported by electro-production data
 corresponding to much lower $q^2$ \cite{Bhattacharya:2011ah}. On the other hand,  this is simply a phenomenological
 fit; there is in principle no  reason why it should work at smaller $q^2$, and other parametrisations  have become recently available.
 %;and there is in principle no  reason why it should work at smaller $q^2$.
 % actually there are indications that the contrary could happen. non ho trovato la ref
 %
 Therefore, for the energy range 
 %task 
 in which we are interested, we lessen the dependence on the dipole approximation  by using 
 %it is better  to use 
simply a linear expansion 
 $g_1/g_1(0)=1+ q^2 r_A^2/6$. The previous value of the axial mass  implies
 $r_A^2=0.455\pm 0.013$ fm$^2$, but a determination that does not assume the double dipole
 has an error larger of about one order of magnitude
  $r_A^2=0.46\pm 0.12$ fm$^2$. 
  
  Proceeding with this conservative estimation, we find 
$\delta\sigma_{\bar\nu_e p}/\sigma_{\bar\nu_e p}=1.1\% (E_\nu/50\mbox{ MeV})^2$ in the 
region above $\sim 10$ MeV; 
   which is actually 
  3 times larger than in \cite{sv}. 
  %The two errors should be summed, and they have the same effect on the cross section at about 15 MeV.

\begin{sidewaystable}
\begin{center}
\begin{tabular}{ ||c|c||c|c||c|c|||c|c||c|c||c|c||   }%|*{3}{*{2}{|l}|}|}
\hline
$E_\nu$  & $\sigma_{\bar\nu_e p}$  & $E_\nu$  & $\sigma_{\bar\nu_e p}$  & $E_\nu$  & $\sigma_{\bar\nu_e p}$ &
$E_\nu$  & $\sigma_{\bar\nu_e p}$  & $E_\nu$  & $\sigma_{\bar\nu_e p}$  & $E_\nu$  & $\sigma_{\bar\nu_e p}$ 
\\{}
 MeV &  10$^{-41}$cm$^2$ & MeV &   10$^{-41}$cm$^2$ &  MeV & 10$^{-41}$cm$^2$ &
 MeV &  10$^{-41}$cm$^2$ & MeV &   10$^{-41}$cm$^2$ &  MeV & 10$^{-41}$cm$^2$
 \\
\hline
$1.9$ & $0.00190183$ & $5.3$ & $0.148898$ & $8.7$ & $0.497688$ &   $2.$ & $0.00331709$ & $35.$ & $8.42244$ & $68.$ & $25.8417$ \\
$2.0$ & $0.00331709$ & $5.4$ & $0.156354$ & $8.8$ & $0.510838$ &  $3.$ & $0.026518$ & $36.$ & $8.86217$ & $69.$ & $26.4317$ \\
$2.1$ & $0.00484224$ & $5.5$ & $0.163984$ & $8.9$ & $0.524148$ &  $4.$ & $0.0680329$ & $37.$ & $9.30948$ & $70.$ & $27.0238$ \\
$2.2$ & $0.00652674$ & $5.6$ & $0.171788$ & $9.0$ & $0.537618$ &  $5.$ & $0.127581$ & $38.$ & $9.76414$ & $71.$ & $27.6179$ \\
$2.3$ & $0.00838532$ & $5.7$ & $0.179765$ & $9.1$ & $0.551247$ &  $6.$ & $0.204734$ & $39.$ & $10.2259$ & $72.$ & $28.2138$ \\
$2.4$ & $0.0104239$ & $5.8$ & $0.187916$ & $9.2$ & $0.565036$ &    $7.$ & $0.299068$ & $40.$ & $10.6946$ & $73.$ & $28.8114$ \\
$2.5$ & $0.0126452$ & $5.9$ & $0.196239$ & $9.3$ & $0.578983$ &    $8.$ & $0.410165$ & $41.$ & $11.1699$ & $74.$ & $29.4107$ \\
$2.6$ & $0.0150505$ & $6.0$ & $0.204734$ & $9.4$ & $0.593089$ &    $9.$ & $0.537618$ & $42.$ & $11.6517$ & $75.$ & $30.0115$ \\
$2.7$ & $0.0176403$ & $6.1$ & $0.213402$ & $9.5$ & $0.607353$ &    $10.$ & $0.681027$ & $43.$ & $12.1398$ & $76.$ & $30.6138$ \\
$2.8$ & $0.0204149$ & $6.2$ & $0.22224$ & $9.6$ & $0.621774$ &      $11.$ & $0.840001$ & $44.$ & $12.6338$ & $77.$ & $31.2174$ \\
$2.9$ & $0.0233742$ & $6.3$ & $0.23125$ & $9.7$ & $0.636352$ &      $12.$ & $1.01415$ & $45.$ & $13.1338$ & $78.$ & $31.8223$ \\
$3.0$ & $0.026518$ & $6.4$ & $0.24043$ & $9.8$ & $0.651088$ &        $13.$ & $1.20311$ & $46.$ & $13.6393$ & $79.$ & $32.4284$ \\
$3.1$ & $0.0298461$ & $6.5$ & $0.24978$ & $9.9$ & $0.665979$ &      $14.$ & $1.4065$ & $47.$ & $14.1504$ & $80.$ & $33.0356$ \\
$3.2$ & $0.0333582$ & $6.6$ & $0.2593$ & $10.0$ & $0.681027$ &      $15.$ & $1.62395$ & $48.$ & $14.6666$ & $81.$ & $33.6438$ \\
$3.3$ & $0.037054$ & $6.7$ & $0.268989$ & $10.1$ & $0.696231$ &    $16.$ & $1.85512$ & $49.$ & $15.188$ & $82.$ & $34.2529$ \\
$3.4$ & $0.040933$ & $6.8$ & $0.278847$ & $10.2$ & $0.711589$ &    $17.$ & $2.09964$ & $50.$ & $15.7143$ & $83.$ & $34.863$ \\
$3.5$ & $0.0449949$ & $6.9$ & $0.288873$ & $10.3$ & $0.727103$ &  $18.$ & $2.35718$ & $51.$ & $16.2452$ & $84.$ & $35.4737$ \\
$3.6$ & $0.0492392$ & $7.0$ & $0.299068$ & $10.4$ & $0.742771$ & $19.$ & $2.6274$ & $52.$ & $16.7808$ & $85.$ & $36.0853$ \\
$3.7$ & $0.0536656$ & $7.1$ & $0.30943$ & $10.5$ & $0.758593$ &   $20.$ & $2.90997$ & $53.$ & $17.3207$ & $86.$ & $36.6974$ \\
$3.8$ & $0.0582736$ & $7.2$ & $0.319959$ & $10.6$ & $0.774569$ & $21.$ & $3.20455$ & $54.$ & $17.8648$ & $87.$ & $37.3101$ \\
$3.9$ & $0.0630629$ & $7.3$ & $0.330655$ & $10.7$ & $0.790698$ & $22.$ & $3.51084$ & $55.$ & $18.413$ & $88.$ & $37.9234$ \\
$4.0$ & $0.0680329$ & $7.4$ & $0.341518$ & $10.8$ & $0.80698$ &   $23.$ & $3.82851$ & $56.$ & $18.9651$ & $89.$ & $38.5371$ \\
$4.1$ & $0.0731833$ & $7.5$ & $0.352546$ & $10.9$ & $0.823414$ & $24.$ & $4.15727$ & $57.$ & $19.5209$ & $90.$ & $39.1511$ \\
$4.2$ & $0.0785136$ & $7.6$ & $0.36374$ & $11.0$ & $0.840001$ &   $25.$ & $4.4968$ & $58.$ & $20.0803$ & $91.$ & $39.7655$ \\
$4.3$ & $0.0840233$ & $7.7$ & $0.3751$ & $11.1$ & $0.85674$ &       $26.$ & $4.84681$ & $59.$ & $20.6432$ & $92.$ & $40.3802$ \\
$4.4$ & $0.0897122$ & $7.8$ & $0.386624$ & $11.2$ & $0.873629$ & $27.$ & $5.20701$ & $60.$ & $21.2093$ & $93.$ & $40.995$ \\
$4.5$ & $0.0955797$ & $7.9$ & $0.398312$ & $11.3$ & $0.89067$ &   $28.$ & $5.57712$ & $61.$ & $21.7787$ & $94.$ & $41.6101$ \\
$4.6$ & $0.101625$ & $8.0$ & $0.410165$ & $11.4$ & $0.907862$ &  $29.$ & $5.95686$ & $62.$ & $22.351$ & $95.$ & $42.2252$ \\
$4.7$ & $0.107849$ & $8.1$ & $0.422181$ & $11.5$ & $0.925204$ &  $30.$ & $6.34595$ & $63.$ & $22.9263$ & $96.$ & $42.8404$ \\
$4.8$ & $0.114249$ & $8.2$ & $0.43436$ & $11.6$ & $0.942696$ &    $31.$ & $6.74412$ & $64.$ & $23.5043$ & $97.$ & $43.4555$ \\
$4.9$ & $0.120827$ & $8.3$ & $0.446702$ & $11.7$ & $0.960337$ & $32.$ & $7.15111$ & $65.$ & $24.085$ & $98.$ & $44.0707$ \\
$5.0$ & $0.127581$ & $8.4$ & $0.459206$ & $11.8$ & $0.978128$ & $33.$ & $7.56666$ & $66.$ & $24.6682$ & $99.$ & $44.6857$ \\
$5.1$ & $0.134511$ & $8.5$ & $0.471872$ & $11.9$ & $0.996067$ &  $34.$ & $7.99052$ & $67.$ & $25.2538$ & $100.$ & $45.3006$ \\
$5.2$ & $0.141617$ & $8.6$ & $0.4847$ & $12.0$ & $1.01415$       &   &   &   &   &   &    \\
\hline
\end{tabular}
\end{center}
\caption{\em Numerical values of the IBD cross section $\bar\nu_e p \to e^+ n$ as a function of neutrino energy by fixing the input parameters at $V_{ud} =0.97427$, $\lambda=1.27601$ and $r_A^2=0.46$ fm$^2$. Left part, low energy region.
 Right part, high energy region.}\label{tab:xsec}
\end{sidewaystable}

\section{Overview}\label{sec:s}
The cross section of the IBD retain its importance for present and future observations.
%for several applications. 
Generally speaking, it seems to be quite well understood. 
Second class currents are not expected to give a significant contribution.
To perform its maintenance for the present needs, 
all we need is a set of consolidated theoretical concepts (that we have thoroughly overviewed)
and, most crucially, we need reliable measurements of the key parameters.

In the range of energies relevant for the detection of  
reactor and supernova electron antineutrinos,  
the cross section depends critically upon $V_{ud}$, $\lambda$ 
and $r_A$.  We have estimated the current 
uncertainties with a conservative procedure. The  uncertainty related to the first two parameters are small and plays a role at low energies; 
it should be added to the one related to the third parameter, which becomes important at higher energies instead. 
When $E_\nu=15$ MeV, the two factors affect the knowledge of the cross section to the same extent.
%Both groups affect the cross section to the same extent  at about 15 MeV.

Note that neutrinos with  different energies are detected in different experiments. 
There is a low energy region that includes
 geoneutrinos (which extend up to about 2.5 MeV) and reactor neutrinos (which end at $\sim$10 MeV);
there is a region of higher energies that includes neutrino fluxes from supernovae (up to 50 MeV) - 
their energy in the interior of the star, during gravitational collapse, is even higher.

How to clarify/improve/progress? We  need to address the reason of discrepancy in   $\tau_n$ measurements.  Even more significant, for the quantitive impact, it is to decrease the uncertainty due to $r_A$--we need to refine the description of the axial form factor in the 100 MeV range.

\paragraph{Acknowledgments:}
We are grateful to Shota Izumiyama for pointing out our inaccuracy in the choice of $r_A^2$.
With partial support of the INFN research initiative ENP and 
of the grant number 
2022E2J4RK,  {\sf\small PANTHEON: Perspectives in Astroparticle and Neutrino
THEory with Old and New Messengers,} part of  PRIN 2022 of the Italian  
{\sf\small Ministero dell'Universit\`a e della Ricerca}.  

\footnotesize

\begin{multicols}{2}

\end{multicols}
\end{document}